\documentclass{svproc}
\newcommand{\bea}{\begin{eqnarray}}
\newcommand{\eea}{\end{eqnarray}}
\usepackage{amsmath}
\usepackage{amsfonts}
\usepackage{amssymb}
\newcommand{\nn}{\nonumber}
\usepackage{url}

\begin{document}
\mainmatter              
\title{Study of the rare decays $B_{s,d}^* \to \mu^+ \mu^-$}
%
%
\author{Suchismita Sahoo$^*$ \and Rukmani Mohanta$^\dagger$}
%
%
%
\institute{School of Physics, University of Hyderabad,
              Hyderabad - 500046, India \\
\email{suchismita@uohyd.ac.in$^*$,}
\email{ rukmani98@gmail.com$^\dagger$}}

\maketitle              

\begin{abstract}
We study  rare leptonic decays of vector $B$ mesons in the scalar leptoquark and family non-universal $Z^\prime$ models. We constrain these new couplings by using the measured branching fractions of $B_{s,d} \to \mu^+ \mu^-$ processes and the existing $B_{s,d}^0-\bar B_{s,d}^0$ mixing data. We estimate the branching fractions of $B_{s,d}^* \to \mu^+ \mu^-$  in both the models, which are found to be reasonably enhanced from their corresponding standard model values and  within the reach of Run 2-3 of LHC.
\end{abstract}
\section{Introduction}
In recent times, the rare $B$ meson decays mediated by  flavour changing neutral current (FCNC) $b \to s,d$ transitions   have been providing  crucial information in our search for new physics (NP) beyond the standard model (SM). The LHCb Collaboration has reported several anomalies in semileptonic $B$ decays at the level of few standard deviations. The rare $B_{s,d} \to \mu^+ \mu^-$ processes are highly suppressed in the SM as they proceed through one loop  and box diagrams. These decays also encounter additional helicity suppression.   Their theory branching fractions are given by \cite{SM}
\bea
&&{\rm BR}(B_s \to \mu^+ \mu^-)|_{\rm SM}=\left (3.65 \pm 0.23 \right ) \times 10^{-9},\nn\\
&&{\rm BR}(B_d \to \mu^+ \mu^-)|_{\rm SM}
=\left (1.06 \pm 0.09  \right ) \times 10^{-10}\;.  \label{brmu}
\eea

In this work, we  investigate the $B_{s,d}^* \to \mu^+ \mu^-$ processes in the context of   leptoquark (LQ) and $Z^\prime$ model.  These processes  don't suffer from  helicity suppression  like $B_{s,d} \to \mu^+ \mu^-$ processes and the only 
 non-perturbative quantity involved is the decay constant of $B_{s,d}^*$ mesons,
which can be precisely calculated from  lattice. LQs are color triplet bosons that couples to both quarks and leptons.   The baryon and lepton number violating LQs  have mass near the grand unification scale to avoid rapid proton decay. Thus we consider the baryon and lepton number conserving LQs, which do not induce proton decay and could be light enough to be accessible in accelerator searches.  On the other hand,  $Z^\prime$  is  a  color singlet  boson, which  could be naturally derived from the extension of  electroweak symmetry of the SM by adding
additional $U(1)^\prime$ gauge symmetry. The family non-universal $Z^\prime$ model \cite{zprime} is the simplest one to explore the inconsistency  between the observed  experimental data and the corresponding  SM
predicted values  in some of the observables associated with $b \to s l^+ l^-$ decays.  The popular “$\pi K$ puzzle” in the hadronic $B \to \pi K$ decays \cite{Pi-K} and other anomalies associated with $b \to s \mu^+ \mu^-$ transitions observed at LHCb could be explained in the $Z^\prime$ model.

The paper is organised as follows. In section 2, we discuss the effective Hamiltonian of $b \to s,d$ transitions in the SM. We then compute the   $B_{s,d}^* \to \mu^+ \mu^-$ processes in the SM as well as LQ  and $Z^\prime$ model. Section 3 contains the conclusion.

\section{$B_{s,d}^* \to \mu^+ \mu^-$ processes}
In the SM, the effective Hamiltonian of rare processes involving the quark-level transitions $b \to q l^+ l^-$, where $q=d,s$ is given by \cite{mohanta2}
\bea
{\cal H}_{\rm eff} = - \frac{G_F}{\sqrt 2} \left [\lambda_t^{(q)}{\cal H}_{\rm eff}^{(t)}
+ \lambda_u^{(q)}{\cal H}_{\rm eff}^{(u) }\right ]+ h.c., \label{ham-sm}
\eea
where
\bea
\mathcal{H}_{\rm eff}^{(u)}&=& C_1(\mathcal{O}_1^c-\mathcal{O}_1^u)+C_2(\mathcal{O}_2^c -\mathcal{O}_2^u), \nn\\
{\cal H}_{\rm eff}^{(t)}&=& C_1 \mathcal{O}_1^c + C_2 \mathcal{O}_2^c + \sum_{i=3}^{10} C_i \mathcal{O}_i,
\eea
$G_F$ is the Fermi constant, $\lambda_k^{(q)}=V_{kb}V_{kq}^*$  and $C_i$'s are the Wilson coefficients.

The decay widths of $B_q^* \to \mu^+ \mu^-$ processes are given by \cite{Grinstein}
\bea\label{br-sm}
\Gamma (B_q^* \to \mu^+ \mu^-) &=& \frac{G_F^2 \alpha^2}{96\pi^3} |V_{tb} V_{tq}^*|^2  f_{B_q^*}^2  m_{B_q^*}^2 \sqrt{m_{B_q^*}^2 -4m_l^2} \times \nn \\ && \left[ \Big{|} C_9^{\rm eff} + 2\frac{m_b}{m_{B_q^*}} C_7^{\rm eff} \Big{|}^2 +  \Big{|} C_{10} \Big{|}^2 \right],
\eea
where $\alpha$ is the fine structure constant, $f_{B_q^*}$ and $m_{B_q^*}$ are the decay constant and mass of $B_q^*$ meson respectively and $m_l$ is the mass of lepton. These  processes are sensitive to the $C_{7,9}^{\rm eff}$ Wilson coefficients, i.e., ${\cal O}_{7}$ and
${\cal O}_{9}$ operators, whereas the contributions
from these operators vanish  in the case of $B_{s,d} \to \mu^+ \mu^-$  processes. In order to calculate the branching fractions, we  need to know the total decay widths of $B_{s, d}^*$ vector bosons.  However, these are
neither measured  nor precisely known   theoretically except for the decay widths of $B_{s,d}^* \to B_{s,d} \gamma$ decays. Now using the particle masses from Ref.  \cite{pdg} and taking the decay widths of radiative $B_{s,d}^*$ bosons from  Ref. \cite{decay-constant}, the branching fractions are found to be  
\bea
{\rm BR}(B_s^* \to \mu^+ \mu^-)|_{\rm SM}&=&\left (1.7 \pm 0.2 \right )  
\left (\frac{0.07~{\rm KeV}}{\Gamma_{B_s^*}^{\rm tot}} \right )
\times 10^{-11},\nn \\
{\rm BR}(B_d^* \to \mu^+ \mu^-)|_{\rm SM}
&=&\left (1.86 \pm 0.21 \right )\left (\frac{0.2~{\rm KeV}}{\Gamma_{B_d^*}^{\rm tot}} \right ) \times 10^{-13}\;.
\eea
These predicted branching fractions are sizable,  about two order lower than the branching fractions of the  corresponding pseudoscalar mesons decay.

 The effective Hamiltonian  in Eq. (\ref{ham-sm}) can be modified in the  both  the LQ and $Z^\prime$ model.  We consider $S_1(3,2,7/6)$ and $S_2(3,2,1/6)$ scalar LQ multiplets, which are invariant under the SM  $SU(3)_C \times SU(2)_L \times U(1)_Y$ gauge group. 
These  will  contribute  new $C_{9,10}^{(\prime) LQ}$  coefficients to the SM. In the NP model, the Wilson coefficients in Eq.  (\ref{br-sm}) will be replaced by $C_{9,10} \to C_{9,10}+C_{9,10}^{(\prime) NP}$. Now comparing the branching fraction of $B_{s,d} \to\mu^+ \mu^-$  with the $1\sigma$ uncertainty of  experimental data, the constraints on the LQ couplings $(\lambda)$ for $M_{LQ}=1$ TeV are found to be \cite{mohanta1}
\bea
 0 \leq \frac{|\lambda^{32} {\lambda^{22}}^*|}{M_S^2} \leq 5 \times 10^{-3},~ ~~
 1.5 \times 10^{-3} ~ \leq \frac{|\lambda^{32} {\lambda^{12}}^*|}{M_S^2} \leq 3.9 \times 10^{-3}~  \;.
 \eea
Using the constrained couplings, we predict the branching fractions of $B_{s,d}^* \to \mu^+ \mu^-$  in both $S_{1,2}$ LQ model and the corresponding values are listed in Table 1.

Similar to the LQ model, the presence of $Z^\prime$ boson  will also provide $C_{9,10}^{Z^\prime}$ new coefficients.  Varying the mass difference of $B_{s,d}^0-\bar B_{s,d}^0$ mixing  within the $2\sigma$ allowed range of experimental value, the constraints on $\rho_q^L$ parameters are  \cite{mohanta2}
\bea
0 \leq \rho_s^L \leq 0.5 \times 10^{-3}~~~1 \times 10^{-4}  \leq \rho_d^L \leq 1.25 \times 10^{-4},
\eea
where $\rho_q^L=\frac{g_2M_Z}{g_1M_Z^\prime}B_{qb}^L$ and $B_{qb}^L$ are the  left handed FCNC $b_L-q_L-Z^\prime$ couplings \cite{zprime2}.  We consider  the coupling of $Z^\prime$ with the leptons as SM like. Using the above constrained couplings, the predicted branching fractions are given in Table 1.   The detailed  calculation of these processes in the SM as well as in both the  LQ and $Z^\prime$ model can be found in Ref. \cite{Grinstein} and \cite{mohanta2}, respectively.
\begin{table}[h]
\caption{Predicted branching fractions  of   $B_{s, d}^* \to l^+ l^-$ decays  in the LQ and  $Z^\prime$ model. }
\begin{center}
\begin{tabular}{| c | c | c| c|}
\hline
 Decay processes   & Values in  & Values in   & Values in \\
&$Y=1/6$ LQ model&$Y=7/6$ LQ model&$Z^\prime$ model\\
 \hline
 
 $B_s^* \rightarrow   \mu^+ \mu^-$ ~ & ~$(1.7-3.19) \times 10^{-11}$~& ~$(1.7-1.93) \times 10^{-11}$~& ~$ (1.7-2.2 ) \times 10^{-11}$~\\

$B_d^* \rightarrow \mu^+ \mu^-$ ~ &  ~$(2.38-8.99)\times 10^{-13}$ ~ &~$(2.47-5.4)\times 10^{-13}$~&~ $(1.67 - 2.23)\times 10^{-13}$ ~\\

 \hline
\end{tabular}
\end{center}
\end{table}

\section{Conclusions}
We have studied the rare leptonic decays of $B_{s,d}^*$ bosons in both the scalar LQ and the family nonuniversal $Z^\prime$ model. We constrain the NP parameters using the measured branching fractions of $B_{s,d} \to \mu^+ \mu^-$ processes and the  $B_{s,d}^0-\bar B_{s,d}^0$ mixing data. We then estimated the branching fractions of $B_{s,d}^* \to \mu^+ \mu^-$ processes, which are found to be sizable and within the reach of LHC experiments.


We thank Science and Engineering Research Board (SERB), Government
of India for financial support through the grant No. SB/S2/HEP-017/2013.
%
%

\end{document}